\documentclass[12pt]{article}
\usepackage[utf8]{inputenc}
\usepackage{geometry}
\usepackage[table,xcdraw]{xcolor}
\usepackage[font={small},labelfont={bf}]{caption}
\usepackage{amsmath}
\usepackage{mathtools}
\usepackage{graphicx}
\usepackage{accents}
\usepackage[colorlinks=true, allcolors=blue]{hyperref}
\usepackage{natbib}
\usepackage{cleveref}
\usepackage{lineno}
\usepackage{authblk}
\usepackage{setspace}

\newcommand{\ie}{i.e., }
\newcommand{\eg}{e.g., }

\makeatletter
\def\@maketitle{%
  \newpage
  \null
  \noindent
  {\Large \bfseries \@title \par}%
  \vskip 1.5em
  {\footnotesize
    \setlength{\parindent}{0pt}
    \setlength{\parskip}{0.3em}
    \@author \par
  }
  \vskip 1.5em
}
\makeatother

\title{A process-based dynamic occupancy model to study range dynamics under non-equilibrium conditions}
\author[1,2]{Simon Lacombe}
\author[2]{Sébastien Devillard}
\author[3]{Cécile Kauffmann}
\author[1]{Olivier Gimenez}

\affil[1]{CEFE, Université de Montpellier, CNRS, EPHE, IRD, 1919 Route de Mende, 34090 Montpellier, France}
\affil[2]{Université Claude Bernard Lyon 1, LBBE, UMR 5558, CNRS, VAS, Villeurbanne, 69622, France}
\affil[3]{SFEPM, 19 allée René Ménard, 18000 Bourges, France}

\begin{document}

\maketitle

\paragraph{Correspondance:}\verb|simon.lacombe@cefe.cnrs.fr|

\doublespacing

\section*{Abstract}
\begin{enumerate}
    \item Many distributions are not at equilibrium with the environment. Failing to account for ecological processes such as dispersal and connectivity when modeling distributions can lead to biased inference about environmental drivers and reduced predictive performance. Spatial dynamic occupancy models are a promising framework to study range dynamics while accounting for dispersal and landscape connectivity, but they currently rely on restrictive formulations of the colonization process, and computational constraints still prevent their application at large spatial scales.
    \item Here, we propose a process-based dynamic occupancy model to study the distribution of range-expanding species while accounting for imperfect detection, connectivity and effects of the environment. We introduce a formulation based on dispersal-pressure that provides a flexible and ecologically interpretable representation of the colonization process, and develop a computational approach based on sparse distance matrices that enables its application to national and transnational scales. 
    \item We conducted a simulation study that showed unbiased parameter estimation across various ecological scenarios We also applied our model to two range-expanding carnivores offering complementary insights: the grey wolf (Canis lupus) and the Eurasian otter (Lutra lutra). Our model revealed contrasting colonization dynamic, with wolves primarily constrained by altitude and forest cover while otters where only marginally affected by the environment, suggesting that their distribution is limited by dispersal history rather than habitat preferences.
    \item By explicitly disentangling the influence of dispersal and environment on distributions, our model provides better insight into occupancy-environment relationships under non-equilibrium conditions, and help identifies what limits species distributions. In light of the increasing availability of large-scale and long-term biodiversity data, our framework offers new opportunities to study range dynamics using mechanistic approaches across entire landscapes.
\end{enumerate}

\vskip 1.5em%
  
{\footnotesize
\leftskip=1cm
\rightskip=1cm
\noindent\textbf{Keywords: \textit{Canis lupus}, connectivity, dispersal, dynamic occupancy models, \textit{Lutra lutra}, process-based models, range-expanding species, species distribution models.}\quad

\par}

\thispagestyle{empty}

\section{Introduction}

Studying species distributions is crucial to assess populations' status and trends, and to offer insights into the drivers of range dynamics, informing conservation strategies. Understanding distribution dynamics also provides a basis for anticipating and managing human-wildlife conflicts, contributing to the development of strategies for coexistence.

Species distributions are dynamic, strongly shaped by historical processes \citep{flojgaard_2009}, species dispersal abilities \citep{sydenham_2017} and landscape dynamics \cite{with_2004}. Failing to account for these elements can lead to flawed inference about environmental drivers of distributions \citep{guisan_2005, broms_2016, guillera-arroita_2017} and strongly limit predictive abilities \citep{yackulic_2015}. Metapopulation ecology aims at explaining how population dynamics and viability is affected not only by the quality of habitat patches, but also by their spatial arrangement and the ability of species to disperse through the landscape \citep{with_2004, howell_2018a}. In that context, the concept of habitat connectivity is defined as the degree to which individuals or propagules can move between suitable habitat patches \citep{coulon_2015} and methods have been developed to study metapopulation dynamics from species occurrence data using models that explicitly include connectivity. Such models have contributed to a better understanding of \eg the interplay between occupancy patterns and dispersal abilities in highly fragmented landscapes \citep{sutherland_2014, dornier_2011} and to provide insights into potential barriers to animal movement \citep{kervellec_2024}. Despite being ever-present in metapopulation studies, concepts such as habitat connectivity and dispersal abilities are rarely modeled explicitly in species distribution models (but see \cite{broms_2016}) where spatial patterns are generally treated as residual spatial autocorrelation instead, and captured using statistical approaches \citep{doser_2022}. While these approaches often allow for reliable inference about effects of the environment and produce realistic distribution maps in most cases \citep{guelat_2018, fidino_2022}, they offer no understanding of the mechanisms responsible for the observed spatial patterns which can range from unaccounted variables to biological processes like dispersal or biotic interactions. Including a mechanistic dimension to model dispersal within distribution models could therefore provide additional knowledge about the drivers of a population. This is particularly valuable for species whose range is limited by factors other than habitat quality, such as landscape barriers to dispersal \citep{kervellec_2024}. In addition, when species distributions are not at equilibrium, occupancy–environment relationships can vary substantially across space and time, leading to biased inference about environmental drivers and reduced predictive performance \citep{yackulic_2015, guillera-arroita_2017}. Such disequilibrium is expected to occur at the edge of a distribution, or when population processes are occurring at a relatively slow pace compared with environmental changes \citep{saltre_2013}, a situation likely to affect an increasing number of species in a rapidly changing world.

Site colonization/extinction models could provide a promising framework to study distributions when equilibrium is not met \citep{yackulic_2015, broms_2016}. In these models, colonization probabilities depend on the occupancy state of surrounding sites, typically through a distance-dependent colonization process \citep{chandler_2015, dornier_2011}. Despite their potential, four limitations have prevented the widespread application of such models. First, they require occurrence data for many sites -- typically pixels over a vast gridded landscape -- across several temporal occasions, which is impossible in most studies due to financial or logistical constraints. Second, even when sampled, imperfect detection causes uncertainty about true occupancy states, affecting inference of colonization and extinction events \citep{guillera-arroita_2017, chandler_2015}. Third, most existing models rely on a similar formulation of the colonization process, often based on restrictive assumptions about how colonization probability responds to the number and proximity of occupied neighboring sites. While this may be of secondary importance in systems with a limited number of isolated sites, in contexts where each site is influenced by numerous contiguous neighbors -- like in distribution studies, this lack of flexibility may affect parameter estimates and limit their ecological interpretability. Finally, colonization-based distribution models require computing connectivity metrics for all pairs of sites, leading to extremely high computational costs making their application virtually impossible for large numbers of sites \citep{kervellec_2024, bled_2011}. To our knowledge, such models have only been used in contexts involving a few dozen \citep{howell_2018a, sutherland_2014} to, in rare cases, around a thousand sites \citep{kervellec_2024, bled_2011}, which is insufficient for large scale distribution studies. For the most part, these first two constraints have been addressed: in the one hand, the development of citizen science programs and community based monitoring has greatly increased the availability of species occurrence data across broad spatial and temporal scales \citep{leandro_2020, louvrier_2018}. On the other hand, advances in occupancy modeling now allow imperfect detection to be explicitly accounted for in a wide range of modeling frameworks \citep{mackenzie_2017}. However, the last two constraints continue to limit the application of site colonization/extinction models to distribution studies. More flexible formulations of the colonization process are needed to provide reliable ecological insights into range dynamics under non-equilibrium conditions, and computational limitations still hinder their application at large spatial scales, motivating the development of new computational approaches.

Here, we propose a process-based dynamic occupancy model to study the distribution of range-expanding species while accounting for non-detection, connectivity and environmental effects. To address the two remaining limitations mentioned above, we introduce a dispersal-pressure formulation that provides a flexible and ecologically interpretable representation of the colonization process, and develop a sparse distance matrix implementation that enables application to large study areas comprising several thousand sites. In practice, this is achieved by defining a maximum dispersal distance a priori, assuming that the influence of distant sites is negligible. We conducted a simulation study to assess whether this approximation allows accurate parameter recovery, and to provide guidelines for validating the choice of the threshold distance. To illustrate the potential of this framework, we applied the model to two recolonizing carnivores at a country-wide scale: the Grey wolf (\textit{Canis lupus}) and the Eurasian otter (\textit{Lutra lutra}) in France. Wolves naturally returned to France in the early 1990s and has since progressively expanded across the country \citep{valiere_2003}. Previous studies used occupancy models to describe this recolonization and identify environmental drivers of wolf distribution, accounting for dispersal indirectly through spatial covariates based on the number of detection in neighboring cells \citep{louvrier_2018}. This system provides an opportunity to assess whether our dispersal-based formulation can recover previously identified patterns while offering a more mechanistic interpretation of colonization dynamics. The Eurasian otter declined across most of Europe during the 20$^{th}$ century but is now recovering following legal protection and improvements in river quality \citep{loy_2022}. Otters have recolonized large areas from several ecologically distinct population refugia, and their current distribution may reflect historical expansion dynamics rather than equilibrium with environmental conditions. The influence of environmental variables on their distribution has often appeared context-dependent \citep{quinonez_2018}. Explicitly accounting for dispersal and historical dynamics may help clarify the ecological drivers underlying otter range expansions. Together, these two case studies offer complementary insights, with wolves offering a well-documented system to assess model behaviour against previous approaches, and otters representing a complex non-equilibrium case in which dispersal processes may strongly confound environmental inference.

\section{Material and Methods}

\subsection{Dispersal-based dynamic occupancy model}

Our model is built on a traditional dynamic occupancy model (\cite{mackenzie_2017}, Equation \ref{eq:dynOccMod}). We consider an array $\mathbf{y}$ of detection/non-detection data across $N$ sites, $T$ primary occasions and $K$ secondary occasions such that $y_{i,t,k} = 1$ if at least one individual was detected at site $i$, primary occasion $t$ and secondary occasion $k$ and $0$ otherwise. Because non-detections can arise either from true absences or from imperfect detection at occupied sites, we define an unknown latent occupancy state $z_{i,t}$ and a detection probability $\rho_{i,t,k}$ and assume occupancy to remain constant across all secondary occasions within a primary occasion, allowing estimation of the detection probability. The occupancy states follows a Markov process, where occupancy at $t+1$ depends on occupancy at $t$ through colonization and extinction processes. Hence, we define for each site an initial occupancy probability $\psi_i$, as well as $\gamma_{i, t}$ and $\omega_{i, t}$, respectively the colonization and extinction probabilities between $t$ and $t+1$.
\begin{equation}
\label{eq:dynOccMod}
    \begin{split}
        y_{i,t,k} &\sim \mbox{Bernoulli}\left(\rho_{i,t,k} \times z_{i,t}\right)\\
        z_{i,t=0} &\sim \mbox{Bernoulli}\left(\psi_{i}\right)\\
        z_{i,t>0} &\sim \mbox{Bernoulli}\left(\gamma_{i, t-1} \times \left(1 - z_{i, t-1} \right) + (1 - \omega_{i,t-1}) \times z_{i,t-1} \right)\\
    \end{split}
\end{equation}
In order to explicitly model the ecological process underlying colonization, we define colonization probability as a function of the state and accessibility of nearby sites on the previous primary occasion. Such models have already been developed in which the colonization probability results from independent colonization attempts as follows: 
\begin{equation}
\gamma_{i,t-1} = 1 - \prod_j\left(1-z_{j,t-1} \times e^{-\frac{d_{i,j}^2}{2\sigma^2}}\right)
\end{equation}
where $e^{-\frac{d_{i,j}^2}{2\sigma^2}}$ is the dispersal kernel, giving the probability that site $i$ is colonized from an occupied site $j$ as a function of distance $d_{i,j}$ and a scale parameter $\sigma$ describing the colonization distance \citep{chandler_2015, kervellec_2024}. Preliminary analyses we conducted revealed that this formulation of the dispersal kernel imposes too strong constraints on the colonization process, preventing the model from disentangling effects of dispersal and environment in some contexts (see Appendix 1 for details). We therefore adopted an alternative formulation inspired by \cite{dornier_2011, sutherland_2014}. We define the dispersal pressure $\Lambda$ such that $\Lambda_{i,t-1}$ is the expected number of dispersers able to reach site $i$ between $t-1$ and $t$. We also define the installation probability $\xi$, as the probability that a site becomes occupied given the arrival of dispersers. We then define the colonization probability for site $i$ between $t-1$ and $t$ as the probability that at least one individual will settle in site $i$ given $\Lambda_{i,t-1}$ and $\xi_i$: 
\begin{equation}
    \gamma_{i,t-1} = \xi_i\times\left(1 - e^{-\Lambda_{i,t-1}}\right)
\end{equation}
The dispersal pressure $\Lambda_{i,t}$ is obtained by summing individual dispersal pressures exerted by all occupied sites $j \neq i$:
\begin{equation}
        \Lambda_{i,t-1} = \frac{\mathcal{A}_i}{2\pi\sigma^2}\times \sum_{j=1}^{N}\left(z_{j, t-1} \times \lambda_je^{-\frac{d_{i,j}^2}{2\sigma^2}}\right)
\end{equation}
 where $\mathcal{A}_i$ is the area of site $i$, and $\lambda_j$ is the dispersal rate from site $j$. The resulting dispersal pressure increases with the number of occupied sites surrounding site $i$, while weighting their contribution by a distance-dependent Gaussian function. As a result, nearby occupied sites contribute strongly to colonization pressure, whereas the influence of distant sites decays at a rate that depends on $\sigma$. In Appendix 1, we show that $\lambda_j$ approximates the expected number of cells colonized from site $j$. To avoid over-parametrization, we assume all sites have the same area $\mathcal{A}$ and dispersal rate $\lambda$. 

 \begin{figure}[htbp]
    \centering
    \includegraphics[width=\linewidth]{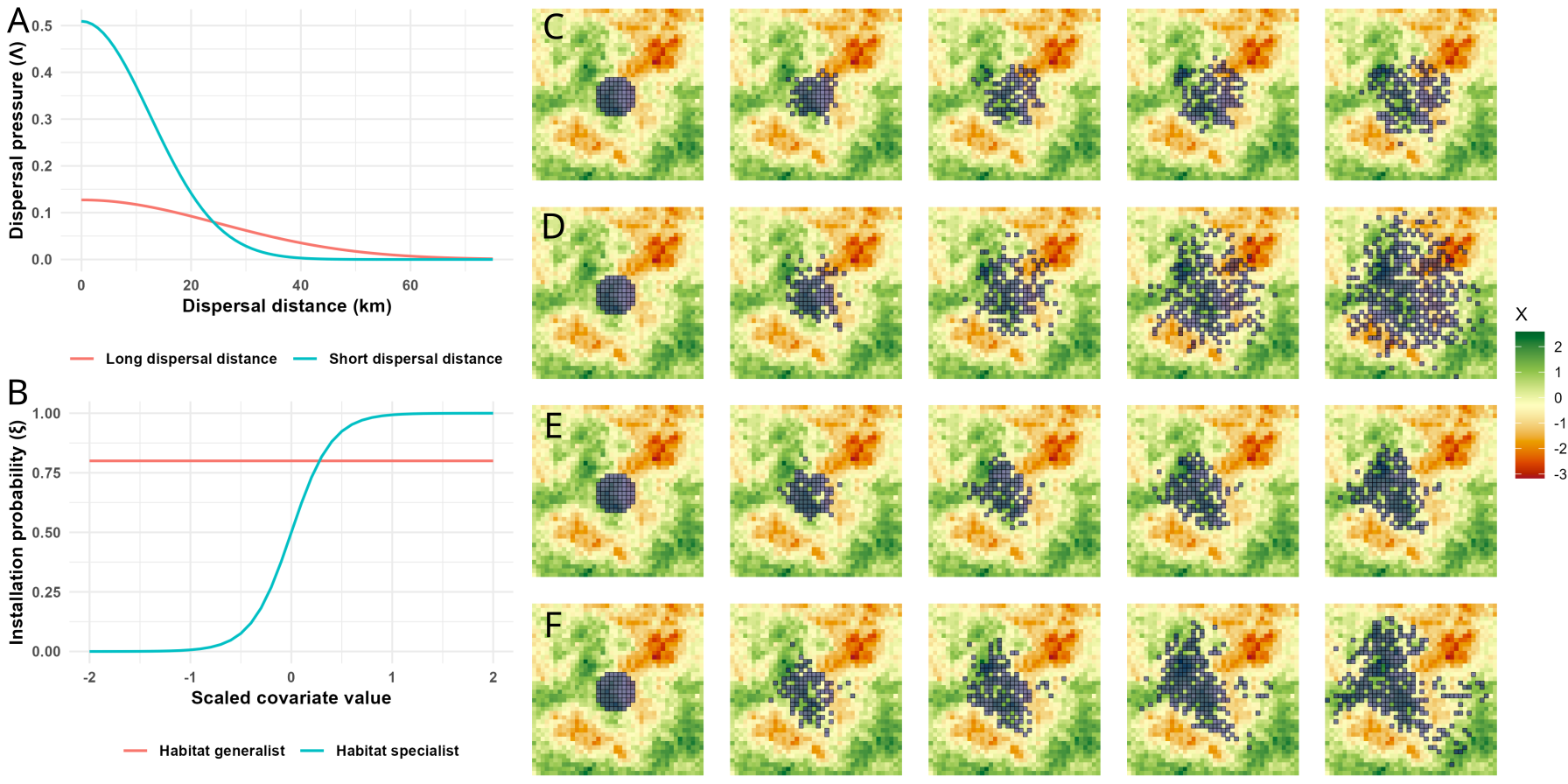}
    \caption{Summary of the four simulation scenarios. (A) dispersal kernel showing the dispersal pressure as a function of the distance to a single occupied cell for theoretical short- and long-distance dispersers. (B) Installation probability as a function of the value of $\mathbf{x}$ for a theoretical habitat generalist and habitat specialist. (C - F) Example of a simulated dataset for each scenario. The value of the covariate $\mathbf{X}$ is shown with background color, with red indicating lowest and green highest values. The latent occupancy state is shown with dark blue cells indicating occupied sites. Panel C. and D. illustrate a generalist species with short- and long-dispersal distance respectively while Panel E. and F. illustrate a specialist species.}
    \label{fig:simulation_scenarios}
\end{figure}

\subsection{Formulation of the model using sparse matrices}

To reduce the computational load of fitting the model over an important number of sites, we defined a threshold distance $d_{max}$ beyond which we assumed the influence of sites on each other negligible. In practice, this was done by retaining only elements of the distance matrix $\mathbf{d}$ verifying $d_{i,j} < d_{max}$, and encoding them using the compressed sparse row (CSR) format, representing the matrix with three vectors:
\begin{itemize}
    \item A \textit{value vector} $\mathbf{v}$ of length $C$ containing all values $d<d_{max}$, ordered by row. 
    \item A \textit{column index vector} $\mathbf{c}$ of length $C$ giving the column indices corresponding to each value in $\mathbf{v}$.
    \item A \textit{row pointer vector} $\mathbf{r}$ of length $N+1$, such that $r_{i<N+1}$ gives the index of the first element of row $i$ in $\mathbf{c}$ and $\mathbf{v}$, and $r_{N+1}=C+1$.
\end{itemize}
Where $C$ is the number of elements of $\mathbf{d}$ satisfying $d_{i,j} < d_{max}$. Hence, all elements $(i,j,d_{i,j})$ of row $i$ are given by the triplets $(i, c_l, v_l)$ for ${l\in\{r_i, ..., r_{i+1}-1\}}$ (see Equation \ref{eq:csr-example} for an example of a dense matrix and its CSR representation).

\begin{equation}
\label{eq:csr-example}
\begin{aligned}
\mathbf{d} &=
\begin{pmatrix}
1 & 0 & 0 & 0 \\
0 & 2 & 0 & 4 \\
0 & 3 & 0 & 0 \\
5 & 0 & 0 & 6
\end{pmatrix}
\\[1em]
\mathbf{v} = (1,2,4,3,5,6), \quad
\mathbf{c} &= (1,2,4,2,1,4), \quad
\mathbf{r} = (1,2,4,5,7)
\end{aligned}
\end{equation}

Finally, we adapted the definition of the propagule pressure to make it compatible with CSR matrices:

\begin{equation}
    \Lambda_{i,t-1} = \frac{\mathcal{A}\lambda}{2\pi\sigma^2}\times \sum_{l=r_i}^{r_{i+1}-1}\left(z_{c_l, t-1} \times e^{-\frac{v_l^2}{2\sigma^2}}\right)
\end{equation}

The CSR format requires storing $2\times C+(N+1)$ values. If each site has on average $\overline{\mathcal{J}}$ neighbors, then $C \simeq \overline{\mathcal{J}} \times N$. The storage complexity therefore scales as $\mathcal{O}(\overline{\mathcal{J}}N)$ which reduces to $\mathcal{O}(N)$ since the size of neighborhoods does not depend on the total number of sites. This is significantly lower than the $\mathcal{O}(N^2)$ complexity of the complete matrix $\mathbf{d}$.

\subsection{Simulation study}

We conducted simulations to assess whether the model could reliably estimate parameter values under different ecological scenarios, and how estimates are affected by setting a maximum colonization distance $d_{max}$. In particular we aimed to provide guidelines regarding the selection of $d_{max}$. We simulated detection/non-detection data across $N = 1600$ sites arranged on a $10$ km grid, $T=5$ primary occasions and $K=4$ secondary occasions, reflecting a realistic large-scale monitoring design. We assumed a constant detection probability $\rho = 0.66$. We set the initial occupancy $z_{i,t=1}$ to 1 in a circle of radius $50$ km located at the center of the study area and 0 elsewhere, corresponding to an initial occupancy $\psi = 0.043$.
We designed the simulations to reflect different types of expanding species. We considered two ecological axes affecting expansion dynamics: dispersal distance and species generalist vs. specialist nature. We set $\lambda = 5$ and used $\sigma = 12.5$ km or $\sigma = 25$ km to represent short and long dispersal distances respectively (Fig. \ref{fig:simulation_scenarios}A). To simulate species generalist vs. specialist nature, we simulated a covariate $\mathbf{X}$ as a Gaussian random field with an exponential covariance function and a scale parameter of $7.5$ km, and defined the installation probability as $\xi = \alpha_\xi + \mathbf{X}\beta_\xi $. We used $\alpha_\xi = logit(0.80)$, $\beta_\xi = 0$ for generalist species, and $\alpha_\xi = 0$, $\beta_\xi = 5$ for specialist species (Fig. \ref{fig:simulation_scenarios}B). Since we were interesting in modeling the colonization process only, we used a constant extinction probability across scenarios of $\omega = 0.20$. We considered four scenarios corresponding to (1) a generalist species with a short dispersal distance (\eg European rabbit \textit{Oryctolagus cuniculus}, Fig. \ref{fig:simulation_scenarios}C), (2) a generalist species with a long dispersal distance (\eg red fox \textit{Vulpes vulpes}, Fig. \ref{fig:simulation_scenarios}D), (3) a specialist species with a short dispersal distance (\eg Hazel dormouse \textit{Muscardinus avellanarius} \ref{fig:simulation_scenarios}E) and (4) a specialist species with a long dispersal distance (\eg Eurasian lynx \textit{Lynx lynx} Fig. \ref{fig:simulation_scenarios}F). 

We simulated 50 datasets per scenario and fitted the dispersal-based occupancy model to each simulated dataset three times, with increasing values of $d_{max}$. We used $d_{max}=25$ km, $37.5$ km and $50$ km for scenarios 1 and 3 and $d_{max}=50$ km, $75$ km and $100$ km for scenarios 2 and 4. This corresponded to $d_{max}=2\sigma$, $3\sigma$ and $4\sigma$ ensuring comparable thresholds between scenarios. For all combinations of scenario and threshold, we computed parameters' relative bias of the mean and coefficient of variation. We also evaluated the proportion of times the 95\% credible interval contained the true value. 

\begin{figure}[hbtbp]
    \centering
    \includegraphics[width=\linewidth]{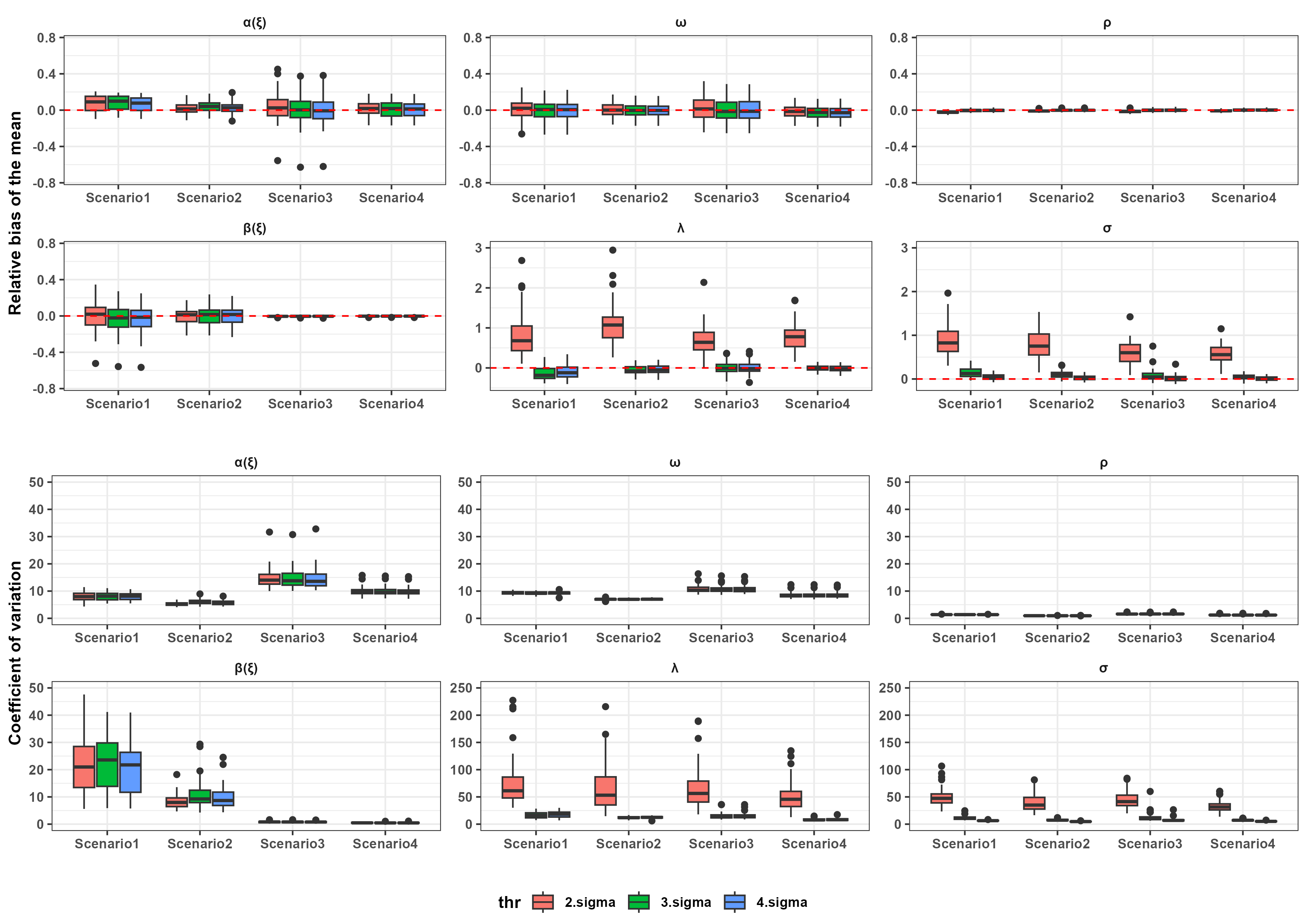}
    \caption{Summary of the simulation study. The first two rows show the relative bias of the posterior mean for each parameter and the last two the coefficient of variation. Panels represent model parameters ($\alpha_\xi$: intercept of the installation probability, $\omega$: extinction probability, $\rho$: detection probability, $\beta_\xi$ slope of the installation probability, $\lambda$ dispersal rate, $\sigma$: scale parameter for the dispersal distance). Scenarios are shown along the x-axis. Colors indicate the threshold distance $d_{max}$ used to fit the model, above which we assume the influence on sites on each other to be null. Note that we use a different y-axis for $\lambda$ and $\sigma$. Boxplots show median values as well as $50\%$ and $95\%$ quantiles.}
    \label{fig:simulation_outputs}
\end{figure}

\subsection{Case study 1: Wolf recolonization in France}

We applied the model to wolf occurrence data collected in France by a network of trained professional and non-professional observers \citep{duchamp_2011} between 1994 and 2016. The data was retrieved from \citep{louvrier_2018} and comprises opportunistic records of scats and footprints monitored each year between December and March. We used the same design as in the original study, defining $N=3450$ 10-by-10 km sites covering a large part of the country, $T=23$ 1-year primary occasions, and $K=4$ secondary occasions corresponding to each month of the monitoring period. In their analysis, the authors constructed a \textit{sampling effort} variable based on the number of people registered in the monitoring network in the vicinity of each site (see details in \cite{louvrier_2018}). We set the detection probability to $0$ in sites where no sampling occurred and used sampling effort and road density to model detection probabilities. Wolves are known to occasionally disperse over very long distances \citep{ciucci_2009}, which can influence range dynamics in ways that our assumption of null dispersal beyond a threshold distance may fail to account for, potentially biasing inferences. To account for potential long distance dispersal we included a background dispersal pressure $\lambda_{min}$, and set $\gamma_{i,t-1} = \xi_i \times \left( 1 - e^{-\left(\Lambda_{i,t-1} + \lambda_{min}\right)}\right)$. We set a $\text{Normal}(\text{logit}(0.05), 1)$ prior to $\lambda_{min}$ on the logit scale to force small values of $\lambda_{min}$ ensuring that the effective dispersal pressure is unaffected near occupied areas, but keeping a small dispersal pressure and therefore a small colonization probability even far from colonization fronts.
We modeled the installation probability as a function of five environmental variables: forest cover, farmland cover, mean altitude, proportion of high altitude, and distance to the closest barrier (highway or major river). See Table S1 for a complete description of all covariates.

\begin{figure}[htbp]
    \centering
    \includegraphics[width=0.5\linewidth]{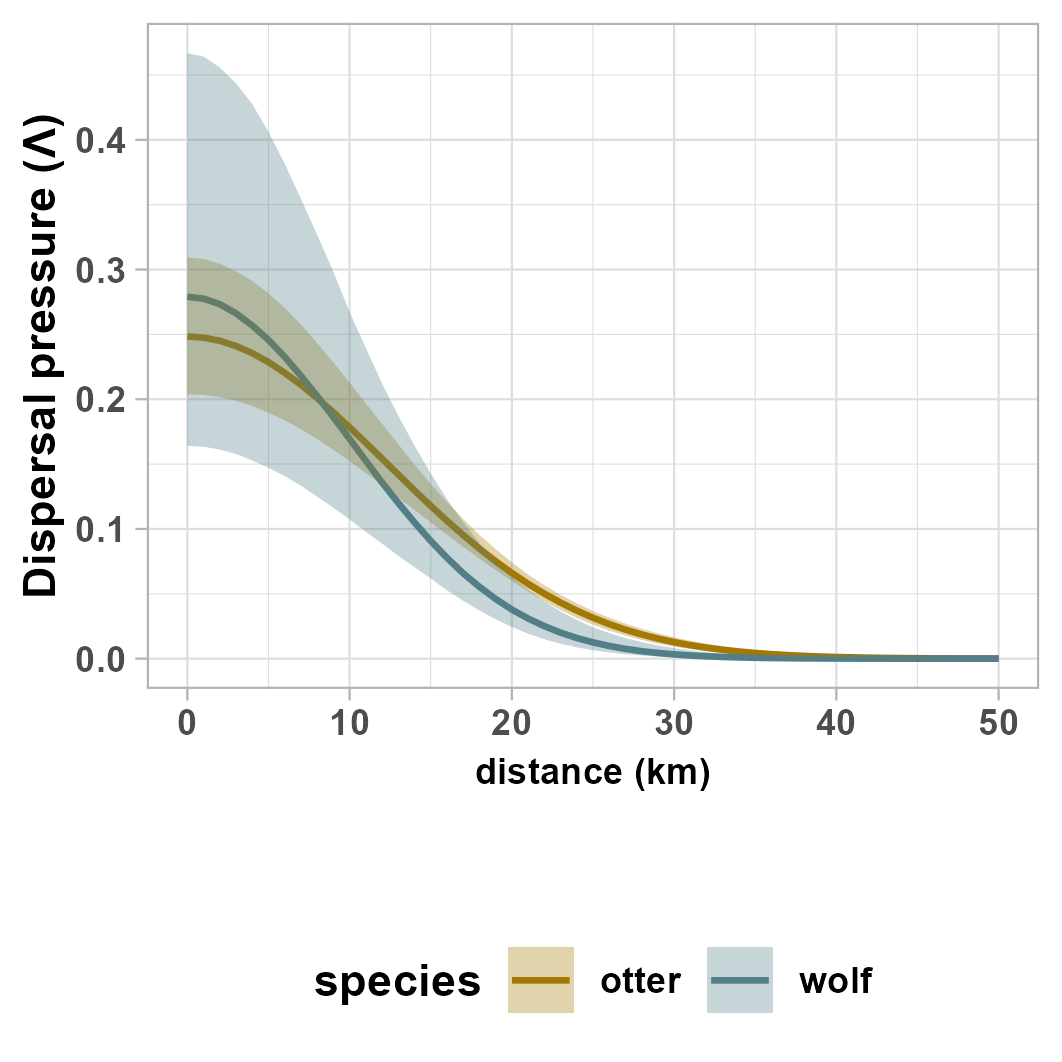}
    \caption{Estimated dispersal pressure exerted by a single occupied cell as a function of the distance for wolves and otters. Solid lines represent median values while shaded ribbons represent $95\%$ credible intervals.}
    \label{fig:disp_kernel}
\end{figure}

\subsection{Case study 2: Otter range expansion in France}

For the second case study, we used a dataset of Eurasian otter occurrences across France between 2000 and 2023. Data was extracted from the Observatoire National des Mammifères data platform (ONM, \url{https://observatoire-mammiferes.fr/}) managed by the French Society for the Study and Protection of Mammals (SFEPM). This platform aggregates otter occurrence data in France collected by naturalist associations, nature users, or professional wildlife managers, and includes both standardized data (mostly based on the IUCN standard protocol \citep{reuther_2000}) and opportunistic observations. We defined our sites as pixels in a 10-by-10 km grid covering the whole mainland France ($N=5478$) to match the resolution at which otter monitoring is conducted across the country \citep{kuhn_2019}. To estimate yearly detection probability, we defined six 4-year primary occasions, treating each year as a secondary occasion, and assuming constant occupancy during these periods. Important heterogeneity in detection probability could arise from the fact that we included both opportunistic presence only data -- prone to sampling biases \citep{backstrom_2024} -- and standardized data. To account for that, we included a \textit{survey} variable indicating whether a standardized survey was conducted in a given site and year. For the $2000-2008$ period, it was defined by extracting negative records from the ONM platform, assuming that surveyed sites provided at least one non-detection during the year. Annual maps of standardized surveys were inspected by experts involved in otter monitoring in France to confirm that they provided reliable information of survey effort. For the $2009-2023$ period, we used the dataset compiled by \citep{lacombe_2025}, in which detection/non-detection data from 24 naturalist associations were consistently classified as standardized surveys or opportunistic observations. We included a linear time trend to account for temporal changes in sampling intensity, estimated separately for sites with and without standardized surveys.

To evaluate effects of the environment on range dynamics, we modeled the installation probability as a function of environmental variables representative of land use, water quality, landscape structure and human disturbances, which have been shown to affect otter presence in certain places. We did not include variables of resource availability as we assume that variation in abundance for different prey groups at a $100$ km² resolution would reflect broad ecological gradients rather than local resource limitation, which is more likely to operate at finer spatial scales. Specifically, we included the proportion of intensive agricultural area, heterogeneous agricultural area, and forest cover in a $25$ m buffer around rivers within each pixel; the diatom biological index of water quality; the cumulated river length and surface of water-bodies, and the terrain slope. See Table S1 for a complete description of all covariates.

\subsection{Implementation in the Bayesian framework}

Models were fitted within the Bayesian framework using Markov Chain Monte Carlo (MCMC). For all analyses, we used $\text{Normal}(0, 2.5)$ priors for the logit intercepts and slopes of $\rho$, $\psi$, $\omega$ and $\xi$, $\text{logNormal}(\text{log}(1), 1)$ and $\text{logNormal}(\text{log}(30), 1)$ for $\lambda$ and $\sigma$ respectively. For the two case-studies we set the maximum dispersal distance to $d_{max} = 50$ km and performed variable selection using the Reversible Jump MCMC algorithm (RJMCMC) \citep{green_1995}. This allows transitions between models including or not each covariate using latent indicator variables, enabling the estimation of inclusion probabilities for all covariates. We ran two MCMC chains with 2500 burn-in iterations and 2500 posterior samples per model in the simulation study and 10000 burn-in iterations and 5000 posterior samples for the two case studies. We assessed convergence for all model by visually inspecting trace plots and computing the Gelman-Rubin $\hat{R}$ statistics \citep{brooks_1998}. All analyses were conducted using R Version 4.4.2 \citep{rcoreteam_2026} and Nimble \citep{deValpine_2017}.

\begin{figure}[ht!]
    \centering
    \includegraphics[width=\linewidth]{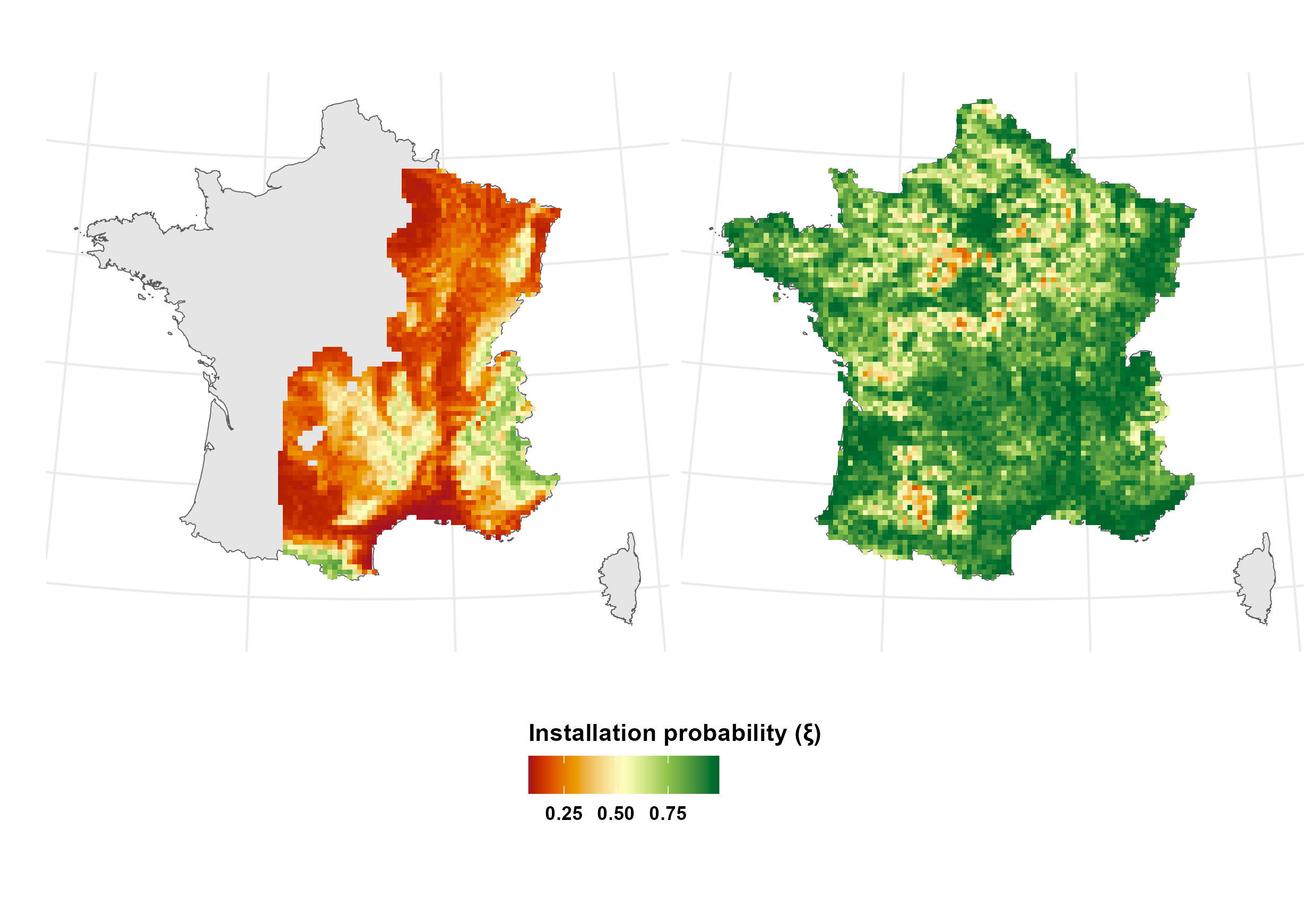}
    \caption{Maps of estimated installation probabilities for wolves (left) and otters (right).}
    \label{fig:ksi_maps}
\end{figure}

\section{Results}

\subsection{Simulation study}
For $\omega$, $\rho$ and $\beta_\xi$, the $50\%$ credible interval of the relative bias overlapped zero for all scenarios, showing unbiased estimation (Fig. \ref{fig:simulation_outputs}A). Parameter $\alpha_\xi$ was slightly overestimated for scenario 1 (short dispersal distance, generalist species) with an estimated relative bias of 0.07 [-0.08, 0.19] (mean [95CI]), but unbiased for other scenarios. Parameters $\lambda$ and $\sigma$ were strongly biased when the model was fitted with $d_{max} = 2\sigma$ with a relative bias of 0.85 [0.24, 2.05] and 0.71 [0.14, 1.40] respectively. This bias decreased markedly with increasing $d_{max}$. For $\lambda$, bias was negligible when $d_{max} \geq 3\sigma$ in all scenarios except the short-distance generalist case, where slight underestimation persisted even at $d_{max} = 4\sigma$ (-0.10 [-0.36, 0.25]). For $\sigma$, unbiased estimation was achieved when $d_{max} = 4\sigma$ for short dispersal distances and when $d_{max} \geq 3\sigma$ for long dispersal distances.

Mean estimated coefficient of variation was consistently below 15\% for $\alpha_\xi$, $\omega$, and $\rho$ (Fig. \ref{fig:simulation_outputs}B) showing good precision. Precision for $\beta_\xi$ was lower in short-distance scenarios, particularly for the short-distance generalist scenario (21.7 [7.62, 40.5]; 9.62 [4.89, 22.6] for scenarios 1 \& 2, respectively) than for long-distance scenarios (0.81 [0.26, 1.49]; 0.50 [0.26, 0.87]). For $\lambda$ and $\sigma$, the coefficient of variation was around 50\% when $d_{max} = 2\sigma$, but decreased to under 20\% when $d_{max} \ge 3\sigma$. 

Overall, when $d_{max} = 4\sigma$ all parameters were estimated with low bias and good coverage, with between 84\% and 100\% of all 95CIs containing the true parameter value (Fig. S3).

\subsection{Case studies}

\begin{figure}[ht!]
    \centering
    \includegraphics[width=\linewidth]{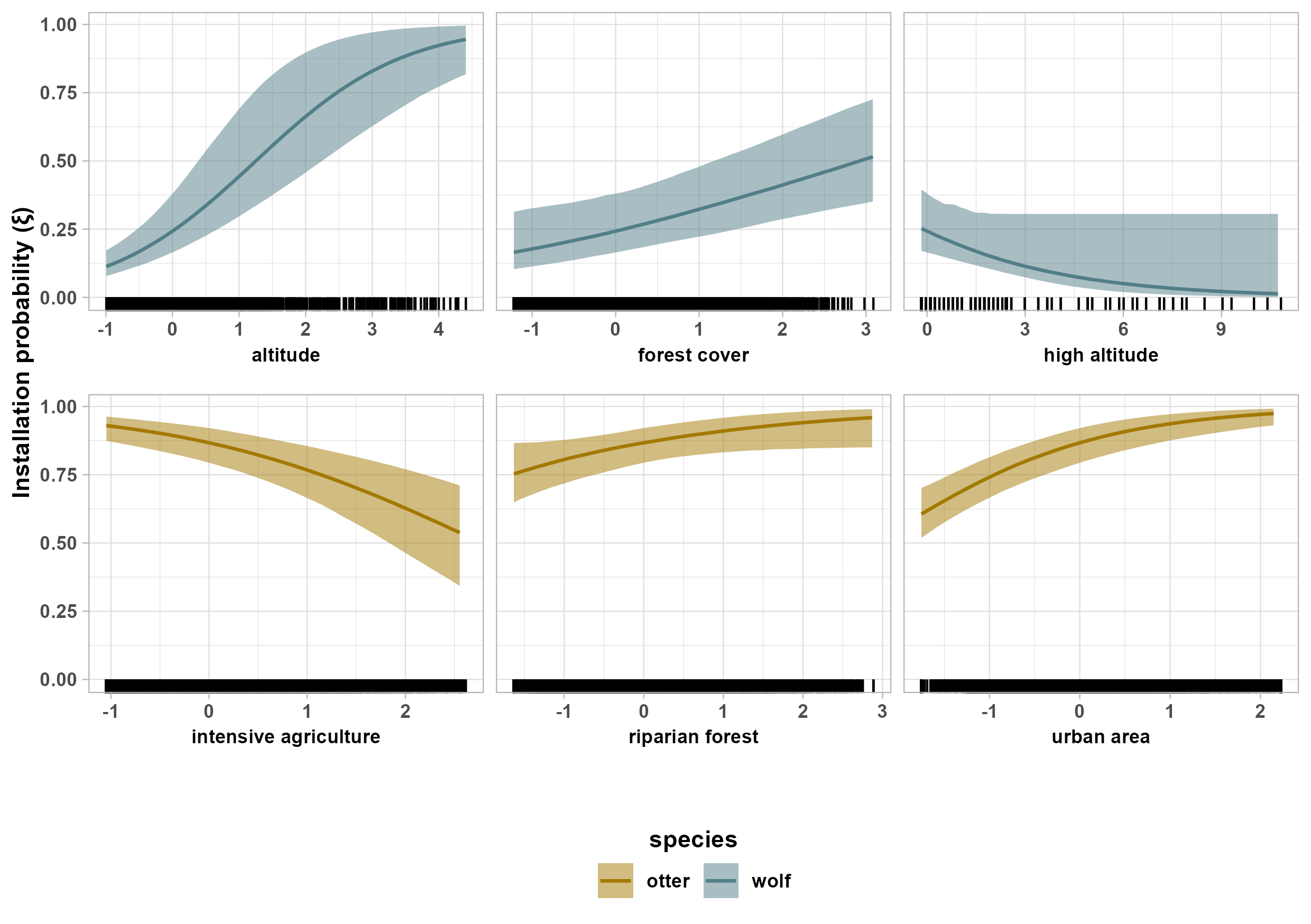}
    \caption{Estimated installation probability ($\xi)$ for wolves (top row) and otters (bottom row) as a function of altitude, proportion of forest cover, proportion of high-altitude, proportion of intensive agricultural area, proportion of riparian forest and proportion of urban area. Only variables with an inclusion probability $>0.66$ are shown. Solid lines represent median values while shaded ribbons represent $95\%$ credible intervals. Vertical segments along the x-axis show covariates' true values for all sites.}
    \label{fig:covs}
\end{figure}

In total, wolves were detected at 2004 1-year primary occasions, with a number of detections ranging from min. 2 sites in 1994 to max. 213 in 2015. When wolves were observed at a site, the average number of positive secondary occasions was 1.80 [1.76, 1.85]. Otters were detected at 9886 4-year primary occasions (min. 695 for 2000-2003, max. 2462 for 2020-2023), with an average number of years with a detection of 2.10 [2.08, 2.13] during positive primary occasions. 

The two models converged, with $\hat{R} \leq 1.05$ for all parameters. All estimates are shown in Table S2. On average, wolves had a low monthly detection probability of 0.14 [0.13, 0.16]. Otter yearly detection probability was higher, reaching 0.72 [0.69, 0.74] where standardized surveys were conducted and 0.40 [0.40, 0.41] elsewhere (Fig. S4). Wolf detection probability was strongly affected by sampling effort and road density, increasing up to 0.69 [0.54, 0.75] where sampling was highest, and to 0.35 [0.32, 0.39] near roads. For otters, yearly detection probability did not vary over time in surveyed sites, but the opportunistic detection probability increased to 0.62 [0.61, 0.63] in 2023. Extinction probability was low for the two species, namely 0.27 [0.25, 0.29] for wolves and 0.14 [0.13, 0.15] for otters. 

Wolves and otters had roughly similar dispersal kernels, with a scale parameters $\sigma$ of 10.72 km [9.15, 12.4] for wolves and 12.26 km [11.4, 13.3] for otters, suggesting a maximum dispersal distance of c.a. 40 km every 4-years for otters and 35 km per year for wolves (Fig. \ref{fig:disp_kernel}). Wolves also had a background dispersal rate of 0.0065 [0.0040, 0.010], reflecting occasional long-distance colonization events. Average installation probability was significantly lower for wolves (0.18 [0.13, 0.26]) than for otters (0.87 [0.80, 0.92]). This resulted in strong differences in maps of installation probability: wolves had overall low installation probabilities, with higher values restricted to a few areas (e.g. Alps, Pyrenean mountains) while otters showed high installation probabilities in most of the study area that decreased in some places to intermediate values (Fig. \ref{fig:ksi_maps}). The RJMCMC algorithm identified several covariates affecting installation probability with high support (inclusion probability $>0.66$, Fig. S5). Altitude had a strong positive effects on wolf installation probability, while forest cover had a slighter positive effect and proportion of high-altitude a negative effect (Fig. \ref{fig:covs}A-C). Otter installation probability was primarily driven by a negative effect of intensive agricultural area, whereas urban areas and riparian vegetation had positive effects (Fig. \ref{fig:covs}D-F).

\section{Discussion}

In this study, we developed a process-based dynamic occupancy model to study the distribution of range expanding species. Our work addresses two key limitations that have so far prevented such models from being applied at large spatial scales. First, we introduced a flexible and ecologically interpretable formulation of the colonization process that allows to simultaneously capture species-specific dispersal patterns and effects of the environment, providing realistic insights into drivers of range dynamics under non-equilibrium conditions \citep{yackulic_2015}. Second, by introducing a computationally efficient approach, we addressed the computational burden associated with connectivity calculations, letting us apply the model to landscapes comprising over 5000 sites, extending the applicability of models initially developed for metapopulations to large-scale distribution studies.

\begin{figure}[htbp]
    \centering
    \includegraphics[width=0.95\linewidth]{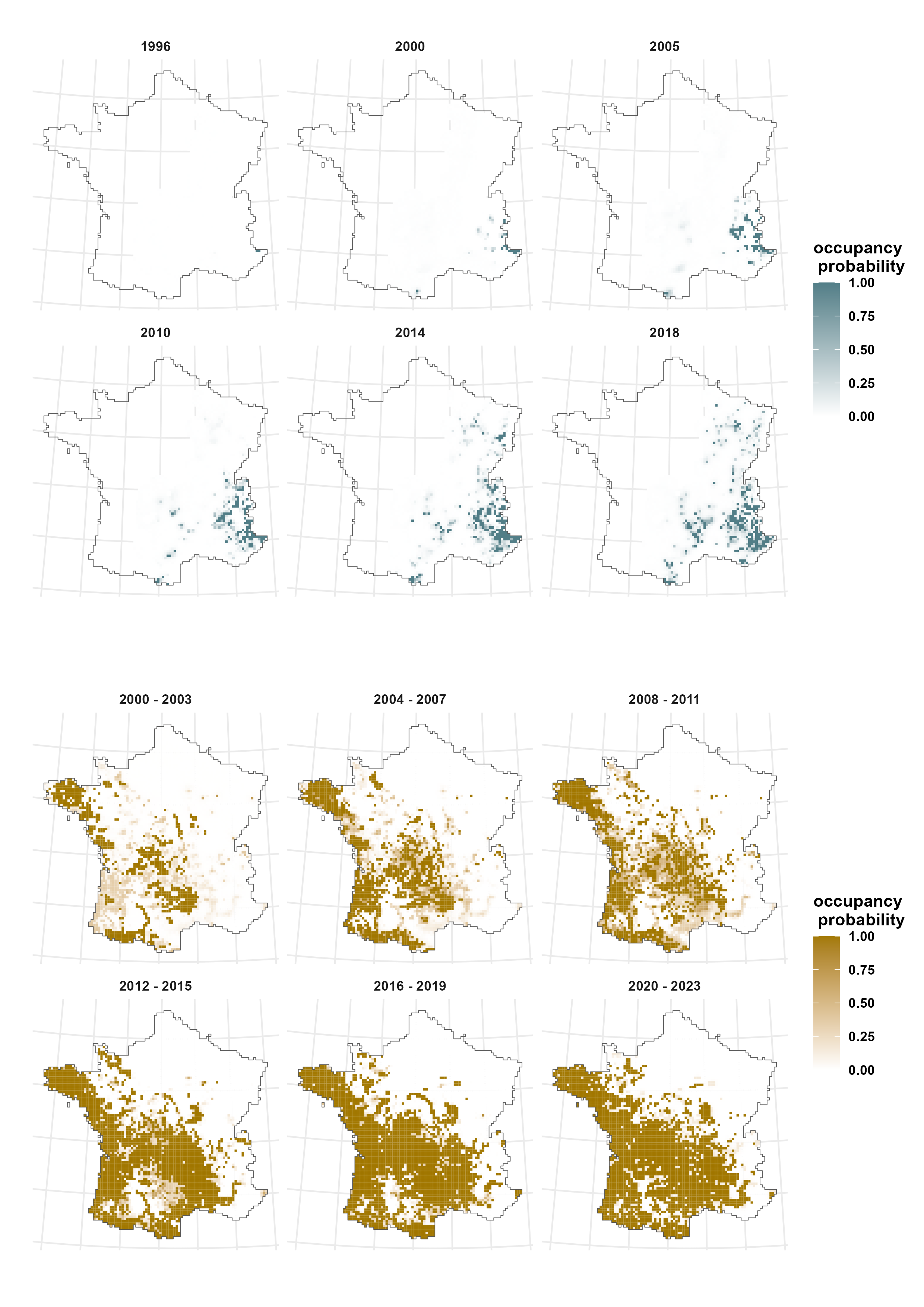}
    \caption{Maps showing the posterior mean of the occupancy probability at each site for the grey wolf (top two rows) and Eurasian otter (bottom two rows).}
    \label{fig:dist_maps}
\end{figure}

In our simulation study, the model could reliably retrieve parameter values for most scenarios when the threshold distance $d_{max}$ was large enough (Fig. \ref{fig:simulation_outputs}). Only for the generalist short-dispersal distance scenario, small biases remained in the posterior mean of the dispersal rate $\lambda$ and the intercept of the installation probability $\alpha_\xi$, showing weak identifiability for these parameters under certain conditions. In that case, it is likely due to the short dispersal distances and high installation probability leading to a near-deterministic progression of the colonization front, causing partial redundancy between parameters. Despite this weak identifiability in specific contexts, we show in Appendix 1 that both parameters are necessary to maintain sufficient flexibility in the dispersal kernel across a broad range of ecological scenarios. This flexibility is critical because overly restrictive dispersal formulations can generate compensatory mechanisms, incorrectly attributing spatial structure to environmental covariates rather than dispersal processes, therefore affecting the model’s ability to disentangle effects of dispersal and environment. 

Too small values of $d_{max}$ would truncate relevant spatial information leading to biased parameter estimates (Fig. \ref{fig:simulation_outputs}). Based on our simulations, we recommend choosing $d_{max}$ to be at least three times the scale parameter of the dispersal process $\sigma$. As $\sigma$ is unknown prior to model fitting, the adequacy of this threshold can only be assessed \textit{a posteriori}, which we consider to be a necessary step before interpreting model estimates. For our case-studies, this assumption was largely met as $\sigma$ was below 12.5 km in both cases.

Our two case-studies revealed contrasting colonization dynamics, despite similar estimated dispersal kernels for otters and wolves (Fig. \ref{fig:disp_kernel}). Wolf installation probability was heterogeneous and strongly affected by the environment (Fig. \ref{fig:ksi_maps}, Fig. \ref{fig:covs}A-C) leading to small and spatially fragmented populations in the Alps and other mountainous areas (Fig. \ref{fig:dist_maps}). The effects of environmental variables were consistent with previous knowledge on this species. In particular, the effects of altitude, forest cover and proportion of high altitude (Fig. \ref{fig:covs}A-C) were similar from those reported in \citep{louvrier_2018}. The persistence of a strong altitude effect after explicitly accounting for dispersal suggests that it reflects an ecological association with mountainous habitats, rather than solely historical colonization patterns \citep{valiere_2003}. This association may be driven by the lower human densities and higher densities of wild ungulates \citep{valiere_2003, loison_2003} in Alpine environments. Interestingly, \cite{louvrier_2018} found a slight effect of farmland cover that our model did not detect. This likely reflects differences in model structure, namely that our model simultaneously estimates dispersal pressure and occupancy states, creating a coupling that may reduce sensitivity to potential biases in detection or sampling effort that could otherwise induce spurious occupancy–environment relationships \citep{risk_2011, guillera-arroita_2017}. By contrast, otter rapidly expanded in all directions resulting in a dense, largely contiguous distribution (Fig. \ref{fig:dist_maps}). In agreement with previous studies, we found otter presence to be negatively associated with cropland surface (see for example \cite{couturier_2023}) and positively with riparian vegetation \citep{looy_2014}. More unexpectedly, we found a positive effects of urban areas. A similar pattern was detected in southern France by \cite{couturier_2023} and likely reflects a combined effect of high survey effort, relatively good quality of freshwater habitats in many peri-urban contexts and otters ability to thrive in heavily urbanized areas \citep{lacombe_2026, weinberger_2016, lee_2025}. Despite this overall consistency, our results diverged from previous studies in several aspects. For instance, we did not find any effects of aquatic habitats (but see \cite{romanowski_2013, quinonez_2018, looy_2014}) or water pollution \citep{couturier_2023}. More generally, although many studies reported strong habitat dependence \citep{romanowski_2013, loy_2009}, environmental effects appeared comparatively moderate in our case (Fig. \ref{fig:ksi_maps}), with a high and relatively homogeneous installation probability across the country. This suggests that, at least in France, part of the observed association between otter presence and environmental variables arises from dispersal history rather than habitat selection. 

The model can be improved by incorporating more functional definitions of connectivity. Indeed, we defined connectivity as a function of Euclidean distance between sites. However, animal capacities to reach distant locations are influenced by their ability to move through complex landscapes \citep{with_2004}. To capture this, Euclidean distance is often replaced in connectivity analyses by functional metrics like least-cost path \citep{coulon_2015, sutherland_2015} or commute-time \citep{mcrae_2008, kervellec_2024} distances, both relying on a representation of the landscape as a resistance surface. Integrating a resistance-based formulation of connectivity would be particularly relevant for species like the otter which primarily disperse along rivers and coastlines \citep{agostini_2025} and where the configuration of watersheds could create functionally isolated habitats despite apparent geographical proximity. More broadly, dispersal is often constrained by anthropogenic or natural barriers that Euclidean distance alone cannot capture. For instance, highways and large rivers have been shown to limit distributions for many large carnivores \citep{kervellec_2023, bauduin_2025}. In the case of wolves, highways reduce movement rates and increase mortality \citep{alexander_2005, ciucci_2009}, potentially affecting range expansion patterns. Although our wolf model partly accounted for such barriers through their effects on installation probability, it did not capture how they affect connectivity. Incorporating functional metrics such as least-cost-path or commute-time distances would therefore represent a major improvement, allowing the model to explicitly assess how landscape features facilitate or impede dispersal. This would not only improve inference on the processes shaping species distributions, but also provide a powerful framework to evaluate how mitigation strategies, such as wildlife crossings, corridor restoration, or barrier removal, may restore connectivity and improve recolonization potential.


A limitation of our approach is that environmental effects are estimated only within the subset of accessible sites, \ie sites the species can access given its estimated dispersal abilities. While this is consistent with the underlying process, it may limit the ability to infer habitat relationships across the full range of environmental conditions within the study area. In particular, because environmental gradients are only partially sampled, the model may fail to capture relationships that occur outside the range of observed conditions, including potential nonlinear responses. This limitation is illustrated by the effect of urbanization on otters, for which the model estimates a positive association, leading to high predicted installation probabilities in highly urbanized areas such as in and around Paris (Fig. \ref{fig:ksi_maps}). However, these areas are unlikely to be suitable for the species \citep{kuhn_2019}, suggesting that the relationship may change at higher levels of urbanization that are not well represented within the accessible portion of the landscape. More generally, this highlights that environmental effects inferred from dispersal-based models should be interpreted with caution, as they may reflect only a subset of the underlying ecological relationships.

To conclude, our model can be interpreted as a process-based null model of range dynamics, in which spatial patterns primarily emerge from species-specific dispersal processes, while environmental effects are estimated as deviations from this baseline. By explicitly disentangling the influence of dispersal and habitat selection, we show that species distributions can reflect fundamentally different dynamics, with some species primarily constrained by the environment and others by their dispersal abilities. These results highlight the need for caution when interpreting habitat–distribution relationships in non-equilibrium systems. In light of the increasing availability of large-scale and long-term biodiversity data and advances in occupancy modeling, our framework offers new opportunities to study range dynamics using mechanistic approaches across entire landscapes. Such approaches provide a promising framework to investigate the processes shaping species distributions under non-equilibrium conditions, while also offering practical tools for anticipating the spread of invasive species, supporting the recovery of threatened populations, and managing human–wildlife interactions.

\paragraph{Acknowledgements}~\\
This research benefited from discussions conducted within the DISCAR working group, funded by the French Foundation for Research on Biodiversity (FRB) through its synthesis center CESAB. 

We would like to thank all volunteer and professional observers who participated in data collection by providing records of otter and wolf occurrence as part of the National Action Plan for the Eurasian Otter and the French Large Carnivore Network. We would also like to thank all regional coordinators involved in the Otter National Action Plan for organizing the standardized surveys in their regions and for sharing their datasets.

We received funding from the City of Montpellier and Montpellier Méditerranée Métropole through the partnership agreement with the Centre for Functional and Evolutionary Ecology, from the University of Montpellier through its Labex Cemeb, and from the Beauval Nature association. This research was partly funded by the ANR through the project nachos for “Interdisciplinary approach to small carnivores - humans relationships” (grant ANR-25-CE03-5469). 

\paragraph{CRediT authorship contribution statement}~\\
\textbf{Simon Lacombe}: Conceptualization, Data curation, Formal analysis, Methodology, Visualization, Writing -- original draft, Writing -- review and editing. \textbf{Sébastien Devillard}: Conceptualization, Supervision, Writing -- review and editing. \textbf{Cécile Kauffmann}: Data curation, Validation, Writing -- review and editing. \textbf{Olivier Gimenez}: Conceptualization, Methodology, Supervision, Writing –- review and editing. 

\paragraph{Declaration of Competing Interest}~\\
The authors declare that they have no known competing financial interests or personal relationships that could have appeared to influence the work reported in this paper.

\paragraph{Declaration of generative AI and AI-assisted technologies in the writing process}~\\
During the preparation of this work, the authors used ChatGPT to polish the text and enhance the English language. After using this tool, the authors reviewed and edited the content as needed and take full responsibility for the content of the published article.

\paragraph{Data availability statement}~\\
Codes and data are available at \url{https://github.com/SimLacombe/PB-DOM}.

\clearpage
\newpage
\footnotesize
\bibliography{references}

@misc{lacombe_2026,
	title = {Comparative efficiency of {eDNA}, camera traps and scat surveys to detect a semi-aquatic mammal across multiple catchments},
	copyright = {© 2026, Posted by openRxiv. This pre-print is available under a Creative Commons License (Attribution-NonCommercial 4.0 International), CC BY-NC 4.0, as described at http://creativecommons.org/licenses/by-nc/4.0/},
	url = {https://www.biorxiv.org/content/10.64898/2026.04.28.721338v1},
	doi = {10.64898/2026.04.28.721338},
	language = {en},
	urldate = {2026-05-05},
	publisher = {bioRxiv},
	author = {Lacombe, Simon and Devillard, Sébastien and D’Hollande, Louise and Raulet, Yann and Sablain, Vincent and Barbu, Louis and Didier, Geoffrey and Mathevet, Raphaël and Miaud, Claude and Oyon, Clément and Pommelet, Eve le and Richarte, Sébastien and Rouvière, Serge and Valentini, Alice and Vazzoler-Antoine, Nathalie and Gimenez, Olivier},
	month = apr,
	year = {2026},
	note = {ISSN: 2692-8205
Pages: 2026.04.28.721338
Section: New Results},
}

@article{alexander_2005,
	title = {Traffic volume and highway permeability for a mammalian community in the {Canadian} {Rocky} {Mountains}},
	volume = {49},
	issn = {1541-0064},
	url = {https://onlinelibrary.wiley.com/doi/abs/10.1111/j.0008-3658.2005.00099.x},
	doi = {10.1111/j.0008-3658.2005.00099.x},
	language = {en},
	number = {4},
	urldate = {2026-04-29},
	journal = {Canadian Geographer / Le géographe canadien},
	author = {Alexander, Shelley M. and Waters, Nigel M. and Paquet, Paul C.},
	year = {2005},
	pages = {321--331},
}

@article{sydenham_2017,
	title = {Disentangling the contributions of dispersal limitation, ecological drift, and ecological filtering to wild bee community assembly},
	volume = {8},
	copyright = {© 2017 Sydenham et al.},
	issn = {2150-8925},
	url = {https://onlinelibrary.wiley.com/doi/abs/10.1002/ecs2.1650},
	doi = {10.1002/ecs2.1650},
	language = {en},
	number = {1},
	urldate = {2026-03-24},
	journal = {Ecosphere},
	author = {Sydenham, Markus A. K. and Moe, Stein R. and Kuhlmann, Michael and Potts, Simon G. and Roberts, Stuart P. M. and Totland, Orjan and Eldegard, Katrine},
	year = {2017},
	pages = {e01650},
}

@article{broms_2016,
	title = {Dynamic occupancy models for explicit colonization processes},
	volume = {97},
	copyright = {© 2016 by the Ecological Society of America},
	issn = {1939-9170},
	url = {https://onlinelibrary.wiley.com/doi/abs/10.1890/15-0416.1},
	doi = {10.1890/15-0416.1},
	language = {en},
	number = {1},
	urldate = {2026-04-27},
	journal = {Ecology},
	author = {Broms, Kristin M. and Hooten, Mevin B. and Johnson, Devin S. and Altwegg, Res and Conquest, Loveday L.},
	year = {2016},
	pages = {194--204},
}

@misc{rcoreteam_2026,
	address = {Vienna, Austria},
	title = {R: {A} {Language} and {Environment} for {Statistical} {Computing}.},
	url = {<https://www.R-project.org/>},
	publisher = {R Foundation for   Statistical Computing},
	author = {{R Core Team}},
	year = {2026},
}

@book{mackenzie_2017,
	title = {Occupancy {Estimation} and {Modeling}: {Inferring} {Patterns} and {Dynamics} of {Species} {Occurrence}},
	isbn = {978-0-12-407245-9},
	shorttitle = {Occupancy {Estimation} and {Modeling}},
	language = {en},
	publisher = {Elsevier},
	author = {MacKenzie, Darryl I. and Nichols, James D. and Royle, J. Andrew and Pollock, Kenneth H. and Bailey, Larissa and Hines, James E.},
	month = nov,
	year = {2017},
}

@article{mcrae_2008,
	title = {Using {Circuit} {Theory} to {Model} {Connectivity} in {Ecology}, {Evolution}, and {Conservation}},
	volume = {89},
	copyright = {© 2008 by the Ecological Society of America},
	issn = {1939-9170},
	url = {https://onlinelibrary.wiley.com/doi/abs/10.1890/07-1861.1},
	doi = {10.1890/07-1861.1},
	language = {en},
	number = {10},
	urldate = {2026-04-21},
	journal = {Ecology},
	author = {McRae, Brad H. and Dickson, Brett G. and Keitt, Timothy H. and Shah, Viral B.},
	year = {2008},
	pages = {2712--2724},
}

@article{yackulic_2015,
	title = {To predict the niche, model colonization and extinction},
	volume = {96},
	copyright = {© 2015 by the Ecological Society of America},
	issn = {1939-9170},
	url = {https://onlinelibrary.wiley.com/doi/abs/10.1890/14-1361.1},
	doi = {10.1890/14-1361.1},
	language = {en},
	number = {1},
	urldate = {2026-03-24},
	journal = {Ecology},
	author = {Yackulic, Charles B. and Nichols, James D. and Reid, Janice and Der, Ricky},
	year = {2015},
	pages = {16--23},
}

@article{doser_2022,
	title = {{spOccupancy}: {An} {R} package for single-species, multi-species, and integrated spatial occupancy models},
	volume = {13},
	issn = {2041-210X},
	shorttitle = {{spOccupancy}},
	url = {https://onlinelibrary.wiley.com/doi/abs/10.1111/2041-210X.13897},
	doi = {10.1111/2041-210X.13897},
	language = {en},
	number = {8},
	urldate = {2023-09-26},
	journal = {Methods in Ecology and Evolution},
	author = {Doser, Jeffrey W. and Finley, Andrew O. and Kéry, Marc and Zipkin, Elise F.},
	year = {2022},
	pages = {1670--1678},
}

@article{chandler_2015,
	title = {Spatial occupancy models for predicting metapopulation dynamics and viability following reintroduction},
	volume = {52},
	copyright = {© 2015 The Authors. Journal of Applied Ecology © 2015 British Ecological Society},
	issn = {1365-2664},
	url = {https://onlinelibrary.wiley.com/doi/abs/10.1111/1365-2664.12481},
	doi = {10.1111/1365-2664.12481},
	language = {en},
	number = {5},
	urldate = {2025-09-24},
	journal = {Journal of Applied Ecology},
	author = {Chandler, Richard B. and Muths, Erin and Sigafus, Brent H. and Schwalbe, Cecil R. and Jarchow, Christopher J. and Hossack, Blake R.},
	year = {2015},
	pages = {1325--1333},
}

@article{guillera-arroita_2017,
	title = {Modelling of species distributions, range dynamics and communities under imperfect detection: advances, challenges and opportunities},
	volume = {40},
	copyright = {© 2016 The Authors},
	issn = {1600-0587},
	shorttitle = {Modelling of species distributions, range dynamics and communities under imperfect detection},
	url = {https://onlinelibrary.wiley.com/doi/abs/10.1111/ecog.02445},
	doi = {10.1111/ecog.02445},
	language = {en},
	number = {2},
	urldate = {2026-03-24},
	journal = {Ecography},
	author = {Guillera-Arroita, Gurutzeta},
	year = {2017},
	pages = {281--295},
}

@article{sutherland_2015,
	title = {Modelling non-{Euclidean} movement and landscape connectivity in highly structured ecological networks},
	volume = {6},
	copyright = {© 2014 The Authors. Methods in Ecology and Evolution © 2014 British Ecological Society},
	issn = {2041-210X},
	url = {https://onlinelibrary.wiley.com/doi/abs/10.1111/2041-210X.12316},
	doi = {10.1111/2041-210X.12316},
	language = {en},
	number = {2},
	urldate = {2026-04-21},
	journal = {Methods in Ecology and Evolution},
	author = {Sutherland, Chris and Fuller, Angela K. and Royle, J. Andrew},
	year = {2015},
	pages = {169--177},
}

@article{valiere_2003,
	title = {Long-distance wolf recolonization of {France} and {Switzerland} inferred from non-invasive genetic sampling over a period of 10 years},
	volume = {6},
	issn = {1469-1795},
	url = {https://onlinelibrary.wiley.com/doi/abs/10.1017/S1367943003003111},
	doi = {10.1017/S1367943003003111},
	language = {en},
	number = {1},
	urldate = {2026-03-20},
	journal = {Animal Conservation},
	author = {Valière, Nathaniel and Fumagalli, Luca and Gielly, Ludovic and Miquel, Christian and Lequette, Benoît and Poulle, Marie-Lazarine and Weber, Jean-Marc and Arlettaz, Raphaël and Taberlet, Pierre},
	year = {2003},
	pages = {83--92},
}

@article{ciucci_2009,
	title = {Long-{Distance} {Dispersal} of a {Rescued} {Wolf} {From} the {Northern} {Apennines} to the {Western} {Alps}},
	volume = {73},
	copyright = {2009 The Wildlife Society},
	issn = {1937-2817},
	url = {https://onlinelibrary.wiley.com/doi/abs/10.2193/2008-510},
	doi = {10.2193/2008-510},
	language = {en},
	number = {8},
	urldate = {2026-03-20},
	journal = {The Journal of Wildlife Management},
	author = {Ciucci, Paolo and Reggioni, Willy and Maiorano, Luigi and Boitani, Luigi},
	year = {2009},
	pages = {1300--1306},
}

@article{fidino_2022,
	title = {Integrated species distribution models reveal spatiotemporal patterns of human–wildlife conflict},
	volume = {32},
	copyright = {© 2022 The Ecological Society of America.},
	issn = {1939-5582},
	url = {https://onlinelibrary.wiley.com/doi/abs/10.1002/eap.2647},
	doi = {10.1002/eap.2647},
	language = {en},
	number = {7},
	urldate = {2023-10-27},
	journal = {Ecological Applications},
	author = {Fidino, Mason and Lehrer, Elizabeth W. and Kay, Cria A. M. and Yarmey, Nicholas T. and Murray, Maureen H. and Fake, Kimberly and Adams, Henry C. and Magle, Seth B.},
	year = {2022},
	pages = {e2647},
}

@article{howell_2018a,
	title = {Increasing connectivity between metapopulation ecology and landscape ecology},
	volume = {99},
	copyright = {© 2018 by the Ecological Society of America},
	issn = {1939-9170},
	url = {https://onlinelibrary.wiley.com/doi/abs/10.1002/ecy.2189},
	doi = {10.1002/ecy.2189},
	language = {en},
	number = {5},
	urldate = {2026-03-20},
	journal = {Ecology},
	author = {Howell, Paige E. and Muths, Erin and Hossack, Blake R. and Sigafus, Brent H. and Chandler, Richard B.},
	year = {2018},
	pages = {1119--1128},
}

@article{flojgaard_2009,
	title = {Ice age distributions of {European} small mammals: insights from species distribution modelling},
	volume = {36},
	copyright = {© 2009 The Authors. Journal compilation © 2009 Blackwell Publishing Ltd},
	issn = {1365-2699},
	shorttitle = {Ice age distributions of {European} small mammals},
	url = {https://onlinelibrary.wiley.com/doi/abs/10.1111/j.1365-2699.2009.02089.x},
	doi = {10.1111/j.1365-2699.2009.02089.x},
	language = {en},
	number = {6},
	urldate = {2026-03-24},
	journal = {Journal of Biogeography},
	author = {Fløjgaard, Camilla and Normand, Signe and Skov, Flemming and Svenning, Jens-Christian},
	year = {2009},
	pages = {1152--1163},
}

@article{bled_2011,
	title = {Hierarchical modeling of an invasive spread: the {Eurasian} {Collared}-{Dove} {Streptopelia} decaocto in the {United} {States}},
	volume = {21},
	issn = {1939-5582},
	shorttitle = {Hierarchical modeling of an invasive spread},
	url = {https://onlinelibrary.wiley.com/doi/abs/10.1890/09-1877.1},
	doi = {10.1890/09-1877.1},
	language = {en},
	number = {1},
	urldate = {2026-03-24},
	journal = {Ecological Applications},
	author = {Bled, Florent and Royle, J. Andrew and Cam, Emmanuelle},
	year = {2011},
	pages = {290--302},
}

@article{backstrom_2024,
	title = {Estimating sampling biases in citizen science datasets},
	issn = {1474-919X},
	url = {https://onlinelibrary.wiley.com/doi/abs/10.1111/ibi.13343},
	doi = {10.1111/ibi.13343},
	language = {en},
	urldate = {2024-08-14},
	journal = {Ibis},
	author = {Backstrom, Louis J. and Callaghan, Corey T. and Worthington, Hannah and Fuller, Richard A. and Johnston, Alison},
	year = {2024},
}

@article{guelat_2018,
	title = {Effects of spatial autocorrelation and imperfect detection on species distribution models},
	volume = {9},
	copyright = {© 2018 The Authors. Methods in Ecology and Evolution © 2018 British Ecological Society},
	issn = {2041-210X},
	url = {https://onlinelibrary.wiley.com/doi/abs/10.1111/2041-210X.12983},
	doi = {10.1111/2041-210X.12983},
	language = {en},
	number = {6},
	urldate = {2026-03-24},
	journal = {Methods in Ecology and Evolution},
	author = {Guélat, Jérôme and Kéry, Marc},
	year = {2018},
	pages = {1614--1625},
}

@article{dornier_2011,
	title = {Colonization and extinction dynamics of an annual plant metapopulation in an urban environment},
	volume = {120},
	copyright = {© 2011 The Authors},
	issn = {1600-0706},
	url = {https://onlinelibrary.wiley.com/doi/abs/10.1111/j.1600-0706.2010.18959.x},
	doi = {10.1111/j.1600-0706.2010.18959.x},
	language = {en},
	number = {8},
	urldate = {2026-03-20},
	journal = {Oikos},
	author = {Dornier, Antoine and Pons, Virginie and Cheptou, Pierre-Olivier},
	year = {2011},
	pages = {1240--1246},
}

@article{saltre_2013,
	title = {Climate or migration: what limited {European} beech post-glacial colonization?},
	volume = {22},
	copyright = {© 2013 John Wiley \& Sons Ltd},
	issn = {1466-8238},
	shorttitle = {Climate or migration},
	url = {https://onlinelibrary.wiley.com/doi/abs/10.1111/geb.12085},
	doi = {10.1111/geb.12085},
	language = {en},
	number = {11},
	urldate = {2026-03-26},
	journal = {Global Ecology and Biogeography},
	author = {Saltré, Frédérik and Saint-Amant, Rémi and Gritti, Emmanuel S. and Brewer, Simon and Gaucherel, Cédric and Davis, Basil A. S. and Chuine, Isabelle},
	year = {2013},
	pages = {1217--1227},
}

@article{kervellec_2024,
	title = {Bringing circuit theory into spatial occupancy models to assess landscape connectivity},
	volume = {15},
	copyright = {© 2024 The Author(s). Methods in Ecology and Evolution published by John Wiley \& Sons Ltd on behalf of British Ecological Society.},
	issn = {2041-210X},
	url = {https://onlinelibrary.wiley.com/doi/abs/10.1111/2041-210X.14418},
	doi = {10.1111/2041-210X.14418},
	language = {en},
	number = {11},
	urldate = {2025-09-12},
	journal = {Methods in Ecology and Evolution},
	author = {Kervellec, Maëlis and Couturier, Thibaut and Bauduin, Sarah and Chenesseau, Delphine and Defos du Rau, Pierre and Drouet-Hoguet, Nolwenn and Duchamp, Christophe and Steinmetz, Julien and Vandel, Jean-Michel and Gimenez, Olivier},
	year = {2024},
	pages = {2141--2152},
}

@article{risk_2011,
	title = {A robust-design formulation of the incidence function model of metapopulation dynamics applied to two species of rails},
	volume = {92},
	copyright = {© 2011 by the Ecological Society of America},
	issn = {1939-9170},
	url = {https://onlinelibrary.wiley.com/doi/abs/10.1890/09-2402.1},
	doi = {10.1890/09-2402.1},
	language = {en},
	number = {2},
	urldate = {2026-04-20},
	journal = {Ecology},
	author = {Risk, Benjamin B. and de Valpine, Perry and Beissinger, Steven R.},
	year = {2011},
	pages = {462--474},
}

@article{sutherland_2014,
	title = {A demographic, spatially explicit patch occupancy model of metapopulation dynamics and persistence},
	volume = {95},
	copyright = {© 2014 by the Ecological Society of America},
	issn = {1939-9170},
	url = {https://onlinelibrary.wiley.com/doi/abs/10.1890/14-0384.1},
	doi = {10.1890/14-0384.1},
	language = {en},
	number = {11},
	urldate = {2026-03-20},
	journal = {Ecology},
	author = {Sutherland, C. S. and Elston, D. A. and Lambin, X.},
	year = {2014},
	pages = {3149--3160},
}

@incollection{loison_2003,
	address = {Berlin, Heidelberg},
	title = {Large {Herbivores} in {European} {Alpine} {Ecosystems}: {Current} {Status} and {Challenges} for the {Future}},
	isbn = {978-3-642-18967-8},
	shorttitle = {Large {Herbivores} in {European} {Alpine} {Ecosystems}},
	url = {https://doi.org/10.1007/978-3-642-18967-8_21},
	doi = {10.1007/978-3-642-18967-8_21},
	language = {en},
	urldate = {2026-04-22},
	booktitle = {Alpine {Biodiversity} in {Europe}},
	publisher = {Springer},
	author = {Loison, A. and Toïgo, C. and Gaillard, J.-M.},
	editor = {Nagy, Laszlo and Grabherr, Georg and Körner, Christian and Thompson, Desmond B. A.},
	year = {2003},
	pages = {351--366},
}

@article{bauduin_2025,
	title = {Modelling {Eurasian} lynx populations in {Western} {Europe}: {What} prospects for the next 50 years?},
	volume = {5},
	issn = {2804-3871},
	shorttitle = {Modelling {Eurasian} lynx populations in {Western} {Europe}},
	url = {https://peercommunityjournal.org/articles/10.24072/pcjournal.543/},
	doi = {10.24072/pcjournal.543},
	language = {en},
	urldate = {2026-04-22},
	journal = {Peer Community Journal},
	author = {Bauduin, Sarah and Germain, Estelle and Zimmermann, Fridolin and Idelberger, Sylvia and Herdtfelder, Micha and Heurich, Marco and Kramer-Schadt, Stephanie and Duchamp, Christophe and Drouet-Hoguet, Nolwenn and Morand, Alain and Blanc, Laetitia and Charbonnel, Anaïs and Gimenez, Olivier},
	month = jun,
	year = {2025},
}

@article{with_2004,
	title = {Metapopulation {Dynamics}},
	url = {https://sci-hub.fr/10.1016/B978-012323448-3/50004-0},
	doi = {10.1016/B978-012323448-3/50004-0},
	urldate = {2026-03-25},
	journal = {Ecology, Genetics and Evolution of Metapopulations},
	author = {With, Kimberly A.},
	year = {2004},
}

@article{guisan_2005,
	title = {Predicting species distribution: {Offering} more than simple habitat models},
	volume = {8},
	issn = {1461-023X},
	shorttitle = {Predicting species distribution},
	doi = {10.1111/j.1461-0248.2005.00792.x},
	journal = {Ecology Letters},
	author = {Guisan, Antoine and Thuiller, Wilfried},
	month = sep,
	year = {2005},
	pages = {993--1009},
}

@article{green_1995,
	title = {Reversible jump {Markov} chain {Monte} {Carlo} computation and {Bayesian} model determination},
	volume = {82},
	issn = {0006-3444},
	url = {https://doi.org/10.1093/biomet/82.4.711},
	doi = {10.1093/biomet/82.4.711},
	number = {4},
	urldate = {2026-03-20},
	journal = {Biometrika},
	author = {Green, P. J.},
	month = dec,
	year = {1995},
	pages = {711--732},
}

@article{duchamp_2011,
	title = {Wolf monitoring in {France} : a dual frame process to survey time- and space-related changes in the population},
	volume = {23},
	issn = {0394-1914, 1825-5272},
	shorttitle = {Wolf monitoring in {France}},
	url = {http://www.italian-journal-of-mammalogy.it/Wolf-monitoring-in-France-a-dual-frame-process-to-survey-time-and-space-related-changes,77271,0,2.html},
	doi = {10.4404/hystrix-23.1-4559},
	language = {english},
	number = {1},
	urldate = {2026-03-20},
	journal = {Hystrix the Italian Journal of Mammalogy},
	publisher = {Associazione Teriologica Italiana},
	author = {Duchamp, Christophe and Boyer, Jérome and Briaudet, Pierre-Emmanuel and Leonard, Yannick and Moris, Perrine and Bataille, Alain and Dahier, Thierry and Delacour, Gilles and Millisher, Gérard and Miquel, Christian and Poillot, Carole and Marboutin, Eric},
	month = jul,
	year = {2011},
	pages = {14--28},
}

@article{brooks_1998,
	title = {General {Methods} for {Monitoring} {Convergence} of {Iterative} {Simulations}},
	volume = {7},
	url = {https://www.tandfonline.com/doi/abs/10.1080/10618600.1998.10474787},
	doi = {https://doi.org/10.1080/10618600.1998.10474787},
	number = {4},
	urldate = {2026-03-20},
	journal = {Journal of Computational and Graphical Statistics},
	author = {Brooks, S. P. and Gelman, .},
	year = {1998},
	pages = {434--455},
}

@article{deValpine_2017,
	title = {Programming with models: writing statistical algorithms for general model structures with \{{NIMBLE}\}},
	volume = {26},
	doi = {10.1080/10618600.2016.1172487},
	number = {2},
	journal = {Journal of Computational and Graphical Statistics},
	author = {de Valpine, P. and Turek, D. and Paciorek, C.J. and Anderson-Bergman, C. and \{Temple Lang\}, D. and Bodik, R.},
	year = {2017},
	pages = {403--417},
}

@article{lee_2025,
	title = {A {Review} on {Eurasian} {Otters} in {Urban} {Areas}: {Principles} for the {Enhancement} of {Biodiversity}},
	volume = {17},
	copyright = {http://creativecommons.org/licenses/by/3.0/},
	issn = {1424-2818},
	shorttitle = {A {Review} on {Eurasian} {Otters} in {Urban} {Areas}},
	url = {https://www.mdpi.com/1424-2818/17/5/356},
	doi = {10.3390/d17050356},
	language = {en},
	number = {5},
	urldate = {2025-10-13},
	journal = {Diversity},
	publisher = {Multidisciplinary Digital Publishing Institute},
	author = {Lee, Connor and Luan, Xiaofeng},
	month = may,
	year = {2025},
	pages = {356},
}

@article{lacombe_2025,
	title = {Range expansion and reconnection of historical populations in the {Eurasian} otter (\textit{{Lutra} lutra}) in {France}: {Insights} from heterogeneous data and integrated species distribution modelling},
	volume = {307},
	issn = {0006-3207},
	shorttitle = {Range expansion and reconnection of historical populations in the {Eurasian} otter (\textit{{Lutra} lutra}) in {France}},
	url = {https://www.sciencedirect.com/science/article/pii/S0006320725002162},
	doi = {10.1016/j.biocon.2025.111179},
	urldate = {2025-08-27},
	journal = {Biological Conservation},
	author = {Lacombe, Simon and Devillard, Sébastien and Kauffmann, Cécile and Aznar, Mélanie and Dupuis, Ondine and Fournier-Chambrillon, Christine and Isère-Laoué, Estelle and Fraissard, Camille and Fuento, Nicolas and Heugas, Tiphaine and Martin, Alexandre and Perrin, Magali and Roche, Antoine and Ruys, Thomas and Simonnet, Franck and Thomas, Bastien and Souriau-Villeger, Angélique and Gimenez, Olivier},
	month = jul,
	year = {2025},
	pages = {111179},
}

@article{leandro_2020,
	title = {Is my sdm good enough? insights from a citizen science dataset in a point process modeling framework},
	volume = {438},
	issn = {0304-3800},
	shorttitle = {Is my sdm good enough?},
	url = {https://www.sciencedirect.com/science/article/pii/S0304380020303537},
	doi = {10.1016/j.ecolmodel.2020.109283},
	urldate = {2025-04-01},
	journal = {Ecological Modelling},
	author = {Leandro, Camila and Jay-Robert, Pierre and Mériguet, Bruno and Houard, Xavier and Renner, Ian W.},
	month = dec,
	year = {2020},
	pages = {109283},
}

@article{agostini_2025,
	title = {A non-invasive genetics insight into population structure and recolonization dynamics of the {Eurasian} otter ({Lutra} lutra) at the boundary of its {Italian} core range},
	issn = {1618-1476},
	url = {https://doi.org/10.1007/s42991-025-00483-1},
	doi = {10.1007/s42991-025-00483-1},
	language = {en},
	urldate = {2025-03-05},
	journal = {Mammalian Biology},
	author = {Agostini, Greta and Loy, Anna and Gentile, Giulia and Giovacchini, Simone and De Sanctis, Cecilia and Mirone, Enrico and Papaleo, Lorenzo and Petrella, Antonio and D’Alessio, Nicola and Colangelo, Paolo},
	month = feb,
	year = {2025},
}

@article{weinberger_2016,
	title = {Flexible habitat selection paves the way for a recovery of otter populations in the {European} {Alps}},
	volume = {199},
	issn = {0006-3207},
	url = {https://www.sciencedirect.com/science/article/pii/S0006320716301537},
	doi = {10.1016/j.biocon.2016.04.017},
	urldate = {2024-11-01},
	journal = {Biological Conservation},
	author = {Weinberger, Irene C. and Muff, Stefanie and de Jongh, Addy and Kranz, Andreas and Bontadina, Fabio},
	month = jul,
	year = {2016},
	pages = {88--95},
}

@article{loy_2009,
	title = {Otter {Lutra} lutra population expansion: assessing habitat suitability and connectivity in southern {Italy}},
	volume = {58},
	language = {en},
	number = {3},
	journal = {Folia Zoologica},
	author = {Loy, Anna and Carranza, Maria Laura and Cianfrani, Carmen and D'Alessandro, Evelina and Marzio, Piera Di and Minotti, Michele},
	year = {2009},
	pages = {309},
}

@article{loy_2022,
	title = {Lutra lutra (amended version of 2021 assessment)},
	shorttitle = {{IUCN} {Red} {List} of {Threatened} {Species}},
	url = {https://www.iucnredlist.org/en},
	urldate = {2024-08-28},
	journal = {The IUCN Red List of Threatened Species 2022: e.T12419A218069689},
	author = {Loy, Anna and Kranz, Andreas and Oleynikov, A. and Roos, A. and Savage, Melissa and Duplaix, Nicole},
	year = {2022},
}

@article{coulon_2015,
	title = {A stochastic movement simulator improves estimates of landscape connectivity},
	volume = {96},
	copyright = {http://onlinelibrary.wiley.com/termsAndConditions\#vor},
	issn = {0012-9658, 1939-9170},
	url = {https://esajournals.onlinelibrary.wiley.com/doi/10.1890/14-1690.1},
	doi = {10.1890/14-1690.1},
	language = {en},
	number = {8},
	urldate = {2024-04-18},
	journal = {Ecology},
	author = {Coulon, A. and Aben, J. and Palmer, S. C. F. and Stevens, V. M. and Callens, T. and Strubbe, D. and Lens, L. and Matthysen, E. and Baguette, M. and Travis, J. M. J.},
	month = aug,
	year = {2015},
	pages = {2203--2213},
}

@article{quinonez_2018,
	title = {A review of otter distribution modeling: {Approach}, scale, and metrics},
	volume = {35},
	shorttitle = {A review of otter distribution modeling},
	journal = {IUCN/SCC Otter Specialist Group Bulletin},
	author = {Quinonez, Ana and Fuller, Todd and Randhir, Timothy},
	month = sep,
	year = {2018},
	pages = {97--127},
}

@article{kervellec_2023,
	title = {Integrating opportunistic and structured non-invasive surveys with spatial capture-recapture models to map connectivity of the {Pyrenean} brown bear population},
	volume = {278},
	issn = {0006-3207},
	url = {https://www.sciencedirect.com/science/article/pii/S0006320722004281},
	doi = {10.1016/j.biocon.2022.109875},
	urldate = {2024-08-14},
	journal = {Biological Conservation},
	author = {Kervellec, Maëlis and Milleret, Cyril and Vanpé, Cécile and Quenette, Pierre-Yves and Sentilles, Jérôme and Palazón, Santiago and Jordana, Ivan Afonso and Jato, Ramón and Elósegui Irurtia, Miguel Mari and Gimenez, Olivier},
	month = feb,
	year = {2023},
	pages = {109875},
}

@book{reuther_2000,
	title = {Surveying and monitoring distribution and population trends of the {Eurasian} otter ({Lutra} lutra) : guidelines and evaluation of the standard method for surveys as recommended by the {European} section of the {IUCN}/{SSC} {Otter} {Specialist} {Group}},
	isbn = {978-3-927650-18-3},
	shorttitle = {Surveying and monitoring distribution and population trends of the {Eurasian} otter ({Lutra} lutra)},
	url = {https://portals.iucn.org/library/node/7962},
	language = {en},
	urldate = {2024-08-15},
	publisher = {GN-Gruppe Naturschutz GmbH, DE},
	author = {Reuther, Claus and Dolch, Dietrich and Green, Rosmary and Jahrl, Jutta and Jefferies, Don and Krekemeyer, Anna and Kucerova, Marcela and Madsen, Aksel Bo and Romanowski, Jerzy and Roche, Kevin and Ruiz-Olmo, Jordi and Teubner, Jens and Trindade, Anabela},
	year = {2000},
}

@article{looy_2014,
	title = {Integrated modelling of functional and structural connectivity of river corridors for {European} otter recovery},
	volume = {273},
	url = {https://hal.science/hal-00933863},
	doi = {10.1016/j.ecolmodel.2013.11.010},
	language = {en},
	urldate = {2023-01-23},
	journal = {Ecological Modelling},
	author = {Looy, K. van and Piffady, Jeremy and Cavillon, C. and Tormos, T. and Landry, P. and Souchon, Y.},
	year = {2014},
	pages = {p. 228},
}

@article{romanowski_2013,
	title = {Habitat correlates of the {Eurasian} otter {Lutra} lutra recolonizing {Central} {Poland}},
	volume = {58},
	issn = {0001-7051},
	url = {https://www.ncbi.nlm.nih.gov/pmc/articles/PMC3606508/},
	doi = {10.1007/s13364-012-0107-8},
	number = {2},
	urldate = {2023-01-21},
	journal = {Acta Theriologica},
	author = {Romanowski, Jerzy and Brzeziński, Marcin and Żmihorski, Michał},
	year = {2013},
	pages = {149--155},
}

@techreport{kuhn_2019,
	address = {Poitiers},
	title = {Plan national d’actions en faveur de la {Loutre} d’{Europe} ({Lutra} lutra) 2019-2028.},
	institution = {SFEPM and DREAL Nouvelle-Aquitaine},
	author = {Kuhn, R. and Simonnet, F. and Arthur, C. and Barthélemy, V.},
	year = {2019},
	pages = {89},
}

@article{couturier_2023,
	title = {Intensive agriculture as the main limiting factor of the otter's return in southwest {France}},
	volume = {279},
	issn = {0006-3207},
	url = {https://www.sciencedirect.com/science/article/pii/S0006320723000277},
	doi = {10.1016/j.biocon.2023.109927},
	urldate = {2023-09-22},
	journal = {Biological Conservation},
	author = {Couturier, Thibaut and Steinmetz, Julien and Defos du Rau, Pierre and Marc, Daniel and Trichet, Emma and Gomes, Régis and Besnard, Aurélien},
	month = mar,
	year = {2023},
	pages = {109927},
}

@article{louvrier_2018,
	title = {Mapping and explaining wolf recolonization in {France} using dynamic occupancy models and opportunistic data},
	volume = {41},
	issn = {09067590},
	url = {https://onlinelibrary.wiley.com/doi/10.1111/ecog.02874},
	doi = {10.1111/ecog.02874},
	language = {en},
	number = {4},
	urldate = {2023-01-23},
	journal = {Ecography},
	author = {Louvrier, Julie and Duchamp, Christophe and Lauret, Valentin and Marboutin, Eric and Cubaynes, Sarah and Choquet, Rémi and Miquel, Christian and Gimenez, Olivier},
	month = apr,
	year = {2018},
	pages = {647--660},
}

\end{document}